\documentclass[twocolumn,superscriptaddress,showpacs,aps,amsmath,amssymb]{revtex4}
\usepackage{bm}
\usepackage{graphicx}
\usepackage{color}

\newcommand{\B}{\text{\scriptsize res}}

\newcommand{\T}{{\rm total}}

\newcommand{\nl}{\nonumber \\}

\newcommand{\ep}{\epsilon}
\newcommand{\w}{\omega}

\newcommand{\be}{\begin{equation}}
\newcommand{\ee}{\end{equation}}
\newcommand{\bea}{\begin{eqnarray}}
\newcommand{\eea}{\end{eqnarray}}
\newcommand{\bsube}{\begin{subequations}}
\newcommand{\esube}{\end{subequations}}

\newcommand{\Fig}[1]{Fig.\,\ref{#1}}

\newcommand{\comments}[1]{}

\begin{document}

\title{Transient Dynamics of a Quantum-Dot in the Mixed Valence Regime}

\author{YongXi~Cheng}
\affiliation{Department of Science, Taiyuan Institute of Technology, Taiyuan
030008, China}

\author{ZhenHua~Li}
\affiliation{ Beijing Computational Science Research Center, Beijing
100084, China }

\author{JianHua~Wei}\email{wjh@ruc.edu.cn}
\affiliation{Department of Physics, Renmin University of China,
Beijing 100872, China }

\author{YiJing~Yan}
\affiliation{Hefei National Laboratory for Physical Sciences at the Microscale,
University of Science and Technology of China, Hefei, Anhui 230026, China}

\begin{abstract}
We investigate the dynamics of a strongly correlated quantum dot system in the mixed valence regime based on the hierarchical equations of motion (HEOM) approach. The transient and steady state transport properties after a quantum quench from equilibration by rapidly applying a bias voltage in a range of temperature below and above the Kondo temperature are described. We find that the time-dependent current exhibits a linear response behavior for weak bias voltage and outside of the linear response regime for larger bias voltage due to the transition of the voltage dependent quantum dot occupancies. The influence of the temperature, finite strongly correlated electron-electron interaction and energy level of the quantum dot on the nonlinear behavior and steady state values of current indicating the Kondo physics are explored in detail.
\end{abstract}


\maketitle
\section{Introduction}
Quantum dots (QDs) as the small regions defined in a semiconductor material with a size of order 100 nm \cite{1993sm118} own the potential applications on quantum computation \cite{2001prb201311} and quantum information \cite{2014nature198}. The wide range of novel physical phenomena of QDs leads to a very active and fruitful research topics, such as artificial atoms, strong Coulomb interaction and coherent time-dependent effects. Especially, the many-body nature of quantum impurity systems can be probed via the QDs devices, such as Kondo effect \cite{2012apl013505}. The investigation of strongly correlated QD systems is helpful to understanding the fascinating collective behavior, such as quantum criticality in heavy fermion systems \cite{2003nature524}, Mott metalinsulator transitions \cite{1998rmp1039}, and high-temperature superconductivity \cite{1986zpb189}. However, the prominent properties of the strongly correlated QD systems is the transient dynamics, both of excited states near the Fermi energy and at highly excited energies \cite{2014prb115139}.

The practical importance of real-time dynamics in QD systems for quantum computing has be emphasized by J. M. Elzerman \emph{et al.}\cite{2004nature431}, the temporal response to gate-voltage pulses for a single-shot is used to detect the spin configuration of a QD in a finite magnetic field \cite{2004nature431,2005prl196801}. Quantum dynamics is discussed in terms of quantum information theory, which indeed facilitates the discussions between physicists, chemists, mathematicians and quantum engineers \cite{2006rpp759}. The real-time dynamics in QD device is of prime importance for our understanding of the quantum dissipation and decoherence and electronic through the nanodevices. More over, the investigation of real-time dynamics in QDs has been successfully used to track individual glycine receptors in the neuronal membrane of living cells \cite{2003science442}.

As a many-body phenomenon, the Kondo effect emerges in the nanoscale Coulomb blockade systems at low temperature. Here, the localized spin and itinerant electrons from reservoirs form a strongly correlated state, which presents a pronounced zero-bias conductance peak at temperatures below the Kondo temperature\cite{1964ptp37,1993cambrige,Ng881768}. In thermal equilibrium, the Kondo problem is well studied and the steady state properties are accurately characterized by a vast amount of analytical and numerical methods, including the many-body perturbation\cite{Kho12075103}, the density matrix renormalization group (DMRG) method \cite{Whi922863,Vid03147902}, the numerical renormalization group (NRG) method\cite{Wil75773,Bul08395} and the quantum Monte Carlo (QMC) approach\cite{Hir862521, Gul11349}, etc.

When a Kondo system is driven out of equilibrium, additional novel Kondo physics appear. Especially, the time dependent Kondo transport problem of QDs devices is still open question. Recent theoretical and experimental efforts aim at observing and modeling nonequilibrium dynamical physics of the Kondo model. A distinct oscillation of the time-dependent current of the one dimensional atomic chain device by applying a bias voltage pulse is presented in terms of nonequilibrium Green function (NEGF) with time domain decomposition(TDD) method. The reason is attributed to the temporal coherence of electrons tunneling through the resonant level in response to the abrupt change of bias\cite{2005prb075317}. The DMRG approach is extended to time-dependent version (TD-DMRG) to explored the time-dependent transport properties for one-dimensional quantum systems and quantum single-impurity system \cite{2002prl256403,2004prl076401,2004jsm04005}. The current-voltage characteristics of the quantum impurity system for mixed valence regime and particle-hole symmetric point are presented \cite{2009prb235336}.  As we known, Wilson's NRG method is a prominent numerical tool for describing the equilibrium Kondo regime \cite{1975rmp773,2008rmp395}. A time-dependent version of NRG(TD-NRG) is developed to investigate the nonequilibrium dynamic of QDs systems \cite{2005prl196801}. Moreover, other numerical methods, such as first principles density functional theory approach\cite{2001prb245407} and Floquet formalism\cite{2000pla419} are also adopted to model the dynamics properties of QDs structure. Except for those attempting works, the perturbative and numerical studies on the transient dynamics through QDs systems are far from extensive due to the computational difficulty and memory effects. For example, the NEGF approach although has been widely adopted in mesoscopic physics, it fails to describe the finite \emph{e-e} interaction case and weakly coupling case\cite{2006prb085324}. The TD-DMRG method is unsuitable for tackling long time scales own to the accumulated error proportional to the time elapsed \cite{2009prb235336}. Furthermore, the time-dependent transport properties in the mixed valence regime of Kondo model which provide some understanding of heavy-fermion compounds has not been studied systematically and a comprehensive picture is missing.

In the present work, we study the real-time dynamics of single QD system by accurately solving the single-impurity Anderson model with the hierarchical equations of motion approach(HEOM) \cite{Jin08234703,2012prl266403}. The geometry is depicted in \Fig{fig1}(a), the QD is in the local magnetic moment regime ($N=1$), and is coupled to the source (L) and drain (R) reservoirs via a coupling strength $\Delta$. The singly-occupied level $\varepsilon_g$ of the single QD is variable modulated by a gate voltage $V_g$. In order to highlight the transient behavior of dynamics in the mixed valence regime, we focus on the time dependent current and occupation following a bias voltage quench $V_{SD}$ from a equilibrium ensemble to a nonequilibrium steady state, as schematically shown in \Fig{fig1}(b). In the Kondo regime, we have reported a temperature dependent oscillation behaviour of the dynamical current through the quantum-impurity system and analyzed the mechanism of time dependent transport oscillations\cite{2015njp033009}. In this paper, as schematically shown in \Fig{fig1}(b), we will characterize the transient behavior of real-time dynamics and time-dependent occupancy response of strongly correlated QD system in mixed valence regime. The results for transient and steady state currents at temperatures ranging from $T \gg T_K$ to $T \ll T_K$ are evaluated. Further more, the effects of finite voltages ($V_{SD}$), strongly correlated electron-electron interaction (\emph{U}) and gate voltage modulating energy level $\varepsilon_{g}$ on time dependent current will be investigated in details. Our results illustrating the time evolution of the current from equilibrium Kondo temperature to the steady state value are relevant for experiments involving potentially important technological applications of QDs and quantum wires.

\section{MODEL AND THEORY}
\begin{figure}
\includegraphics[width=0.85\columnwidth]{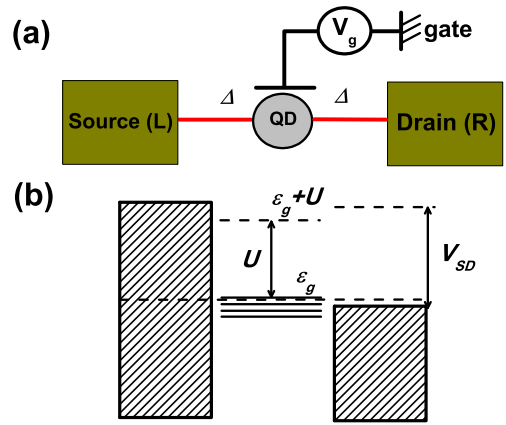}
\caption{(a) Schematic representation of the single QD device. The QD is embedded between two leads (Source and Drain) via the dot-lead coupling strength $\Delta$. The gate voltage $V_{g}$ controls the dot charge. (b) The dynamical transport exhibits with the system subject to an step voltage with constant value $V_{SD}$. $\varepsilon_{g}$ and $U$ are the energy level and the coulomb interaction at the QD, respectively.}
\label{fig1}
\end{figure}
In the geometry depicted in figure \ref{fig1} (a), a QD is directly coupled to the source (L) and drain (R) via hybridization widths $\Delta$, and the QD' energy is tunable via the gate voltage $V_{g}$. A simple total Hamiltonian able to describe the system is $H=H_{dot}+H_{leads}+H_{coupling}$, where the isolated QD is described by the single-impurity Anderson model
\begin{align}\label{hd}
   H_{dot}=\sum_{\sigma}(\varepsilon_{g}+eV_g)\hat{a}^\dag_{\sigma}\hat{a}_{\sigma} + U n_{\sigma}n_{\bar{\sigma}}
  \end{align}
here, $\hat{a}_{\sigma}^\dag$ ($\hat{a}_{\sigma}$) corresponds to the operator that creates (annihilates) a spin-$\sigma$ electron with energy $\epsilon_{g}$ in the QD. Gate voltage $V_{g}$ controls the energy level $\epsilon_{g}$ via modulate the dot charge. $n_{\sigma}=\hat{a}^\dag_{\sigma}\hat{a}_{\sigma}$ corresponds to the operator of the electron occupation number at the QD. $U$ is on-dot Coulomb interaction between electrons with spin $\sigma$ and $\bar{\sigma}$ (opposite spin of $\sigma$).

The Hamiltonian of source and drain is
\begin{align}\label{hl}
   H_{leads}=\sum_{k\mu\alpha=L,R}(\epsilon_{k\alpha} + \frac{\alpha V_{SD}(t)}{2})\hat{d}^\dag_{k\mu\alpha}\hat{d}_{k\mu\alpha}
  \end{align}
with $\hat{d}^\dag_{k\alpha}$($\hat{d}_{k\alpha}$) corresponding the creation (annihilation) operator for an electron with the $\alpha$-reservoir state $|k\rangle$ of energy $\epsilon_{k\alpha}$. $\epsilon_{k\alpha}$ is the energy of an electron with wave vector $k$ in the $\alpha$ lead. $V_{SD}(t)$ is the time dependent voltage of the two leads.

The dot-lead coupling part is
\begin{align}\label{hc}
H_{coupling}=\sum_{k\mu\alpha}t_{k\mu\alpha}\hat{a}^\dag_{\sigma}\hat{d}_{k\mu\alpha}+\text{H.c.}
  \end{align}

To describe the stochastic nature of the transfer coupling, the dot-lead coupling part can be written in the reservoir $H_{\rm leads}$-interaction picture as $H_{coupling}=\sum_{\mu}[f^\dag_{\mu}(t)\hat{a}_{\mu} +\hat{a}^\dag_{\mu}f_{\mu}(t)]$. Where $f^\dag_{\mu}=e^{iH_{leads}t}[\sum_{k\alpha}t^{*}_{k\mu\alpha}\hat{d}^\dag_{k\mu\alpha}]e^{-iH_{leads}t}$ is the stochastic interactional operator and satisfies the Gauss statistics, with $t_{k\mu\alpha}$ denoting the transfer coupling matrix element. Therefore, the influence of electron reservoirs (source and drain) on the QD can be characterized by the hybridization bath spectral density functions $\Delta_{\alpha}(\w)\equiv\pi\sum_{k} t_{\alpha k\mu}t^\ast_{\alpha k\mu} \delta(\w-\ep_{k\alpha})=\Delta W^{2}/[2(\w-\mu_{\alpha})^{2}+W^{2}]$, here $\Delta$ is the effective QD-lead couplings, $W$ is the band width of reservoirs, and $\mu_{\alpha}$ is the chemical potentials of the $\alpha$ reservoir \cite{2013zheng086601,2015njp033009,2012prl266403}.

In this paper, the single-impurity Anderson model is accurately solved by the hierarchical equations of motion (HEOM) approach. The HEOM established based on the Feynman-Vernon path-integral formalism is potentially useful for treating the quantum-impurity systems, here the system-bath correlations are fully take into consideration \cite{2013zheng086601,2012prl266403,2017prb155417}. This formalism is also valuable for addressing both static and transient electronic properties of strongly correlated system. In principle, the HEOM formalism is formally rigorous for noninteracting electron reservoirs, and the numerical results of HEOM are quantitatively accurate, as the date converge uniformly with respect to the truncation of the hierarchy \cite{2016advanced}. The outstanding issue of characterizing both equilibrium and nonequilibrium properties of open quantum systems are referred to Refs. \cite{Jin08234703,2012prl266403,2016advanced}. Especially, The HEOM approach has been employed to study transient dynamic properties of strongly correlated quantum-impurity systems, including the time dependent transport properties \cite{2015njp033009}, dynamic Kondo memory phenomena \cite{2013zheng086601} and dynamic Coulomb blockade \cite{2008njp093016,2009jcp164708}. In ref. \cite{2015njp033009} we have compared the results obtained from HEOM approach to  other methods, it has been demonstrated that the HEOM can achieve the same level of accuracy as the latest NRG method, DMRG approach and QMC approach \cite{2012prl266403,2015njp033009}.

The reduced density matrix of the QDs system $\rho^{(0)}(t) \equiv {\rm tr}_{\B}\,\rho_{\T}(t)$ and a set of auxiliary density matrices  $\{\rho^{(n)}_{j_1\cdots j_n}(t); n=1,\cdots,L\}$ are the basic variables of HEOM approach. Here, $L$ denotes the terminal or truncated tier level. The numerical results of HEOM program can converge uniformly with truncation $L$. The equations that governs the dynamics of open system assumes the form of \cite{Jin08234703,2012prl266403}:
\begin{align}\label{HEOM}
   \dot\rho^{(n)}_{j_1\cdots j_n} =& -\Big(i{\cal L} + \sum_{r=1}^n \gamma_{j_r}\Big)\rho^{(n)}_{j_1\cdots j_n}
     -i \sum_{j}\!     
     {\cal A}_{\bar j}\, \rho^{(n+1)}_{j_1\cdots j_nj}
\nl &
    -i \sum_{r=1}^{n}(-)^{n-r}\, {\cal C}_{j_r}\,
     \rho^{(n-1)}_{j_1\cdots j_{r-1}j_{r+1}\cdots j_n},
\end{align}
with ${\cal A}_{\bar j}$ and ${\cal C}_{j_r}$ being the Grassmannian superoperators \cite{Jin08234703,2012prl266403}.

The varied physical quantities can be acquired via the HEOM-space linear response theory \cite{Wei2011arxiv}. We prepare the initial total system at equilibrium, where $\mu_{\alpha}=\mu^{eq}=0$. When the system contain time-dependent voltage $V_{SD}(t)$, the system out of equilibrium, and the resulting time evolutions of the QD occupancy is
\begin{align}\label{hd}
    n_{\mu}(t)=\sum_{\mu} n_{\mu}(t)=tr_{s}[\hat{a}^\dag_{\mu}\hat{a}_{\mu}\rho^\dag_{\alpha \mu}(t)]
  \end{align}

The electric current from $\alpha$-lead to the single QD system is given by
\begin{align}\label{hd}
    I_{\alpha}(t)=i\sum_{\mu}\mathrm{tr}_{s}[{\rho^\dag_{\alpha \mu}(t)\hat a_{\mu} -\hat a^\dag_{\mu}\rho^-_{\alpha \mu}(t)}],
  \end{align}
with $\rho^\dag_{\alpha \mu}=(\rho^-_{\alpha \mu})^\dag$ being the first-tie auxiliary density operator. The details of the HEOM formalism and the derivation of physical quantities are supplied in the Refs.~\cite{Jin08234703,2012prl266403}.

\section{RESULTS AND DISCUSSION}

We present the numerical results for the transient dynamics of the single QD model in the mixed valence regime $\varepsilon_{g} = 0$ using the HEOM approach. To compare conveniently to experiment, the parameters adopted are similar to Refs ~\cite{2013na145,2016prl036801}. The electron-electron interaction of the QD is assumed $U = 3\mathrm{meV}$. The QD-lead coupling strength is $\Delta = 0.3\mathrm{meV}$, the band width of leads is $W = 5\mathrm{meV}$. The Kondo temperature $T_K$ is on the order of $\Delta$  and the system is equilibrate with times $\propto \Delta^{-1}$ \cite{2016prl036801}. Firstly, we focus on the current-voltage characteristics. Here, we adopt an low temperature as $K_{B}T = 0.03\mathrm{meV}$. Figure \ref{fig1} shows the evolution of the time dependent current (Fig. \ref{fig1}(a)) and QD occupancy (Fig. \ref{fig1}(b)) of the single QD system subject
to various forms of the time-dependent step voltage.
\begin{align}\label{hd}
   V(t)=\left\{
   \begin{array}{ll}
   0 \;\,\qquad(t<0)& \\
   V_{SD} \qquad(t\geq 0) &
   \end{array}
    \right.
  \end{align}%
$V_{SD}$ is the voltage quench, adopted a range of values between $0.1\mathrm{meV}$ and $1.0\mathrm{meV}$. When the voltage quench applied to the source and drain, the current flowing through the device
engenders. We observe that for all voltage pulses, the current equilibrates at time $t \approx 5 \mathrm{ps}$ and the steady state values of current increase with the applied voltages. It is interesting to see that the time-dependent current exhibits a linear response behavior for weak bias voltage and outside of the linear response regime for larger bias voltage. For the low bias voltages (such as $V_{SD} = 0.1\mathrm{meV}$ and $0.3\mathrm{meV}$ ), the current shows a monotonic rise and saturation to a steady state value. For the strong bias voltages ($V_{SD} > 0.5\mathrm{meV}$), After the current rapidly increases to a maximal value, nonlinear behavior emerge. The linear response behavior only at small times ($t < 1.0\mathrm{ps}$), and the nonlinear behavior is visible for a larger time ($1.0\mathrm{ps} < t < 5.0\mathrm{ps}$). Additionally, the nonlinear behavior depends strongly on the size of the applied step voltage. With increasing the bias voltages, the nonlinear behavior engenders early and becomes oscillations of current. For example, the nonlinear behavior engenders at $t = 1.5\mathrm{ps}$ for the voltage $V_{SD}=0.50\mathrm{mV}$, while it emerges at $t = 1.0\mathrm{ps}$ for the voltage $V_{SD}=1.0\mathrm{mV}$, accompanied by increasing current oscillations. This illustrates the breakdown of conductance at large voltages. The corresponding QD occupancies are characterized in figure \ref{fig1}(b). It is clearly visible that the occupancies increase with the voltage pulse, which demonstrates the stronger voltage dependent nonlinear behavior of the transport. Moreover, both the steady state values of current and of dot occupancy at lager times scale with the voltage quench.
\begin{figure}
\includegraphics[width=0.95\columnwidth]{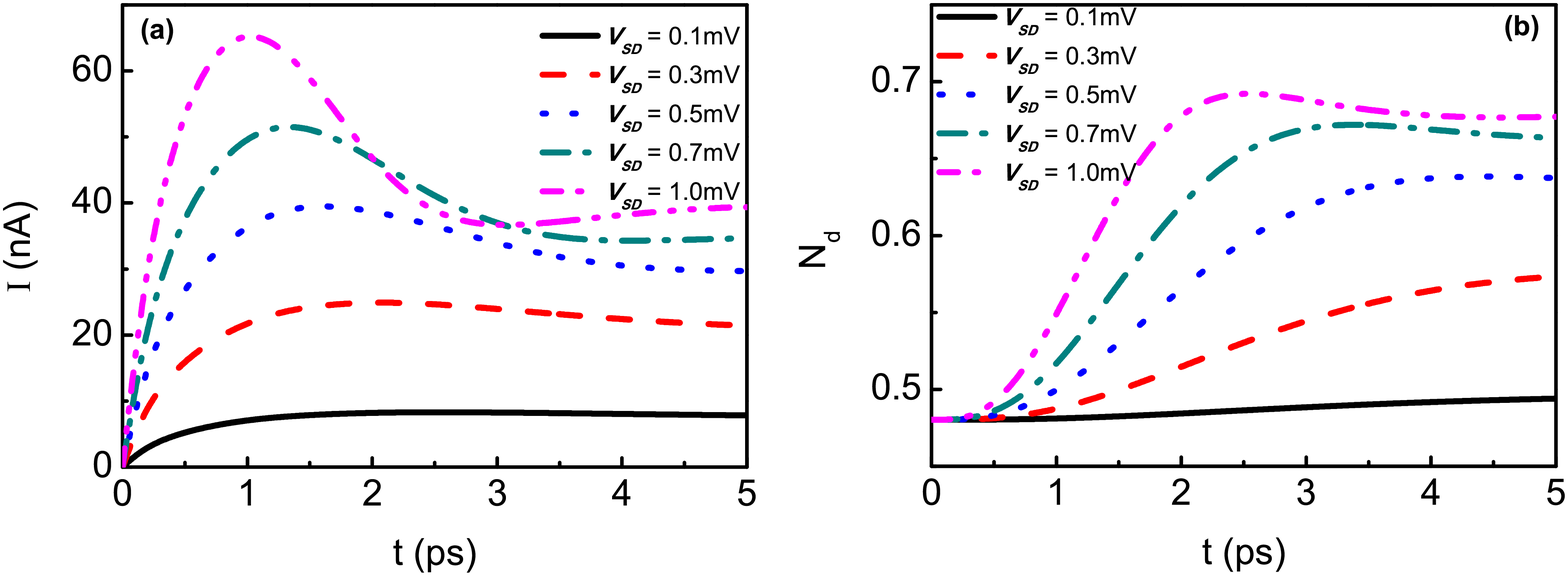}
\caption{(Color online). (a) Current $I(t)$ as a function of time $t$ for different voltages $V_{SD}$. (b) Dot occupancy as a function of time $t$ for different voltages $V_{SD}$. The other parameters are
$\varepsilon_{g} = 0\mathrm{meV}$, $U = 3\mathrm{meV}$, $W = 5\mathrm{meV}$, $\Delta = 0.3\mathrm{meV}$ and $K_{B}T = 0.03\mathrm{meV}$.}
\label{fig2}
\end{figure}

\begin{figure}
\includegraphics[width=0.95\columnwidth]{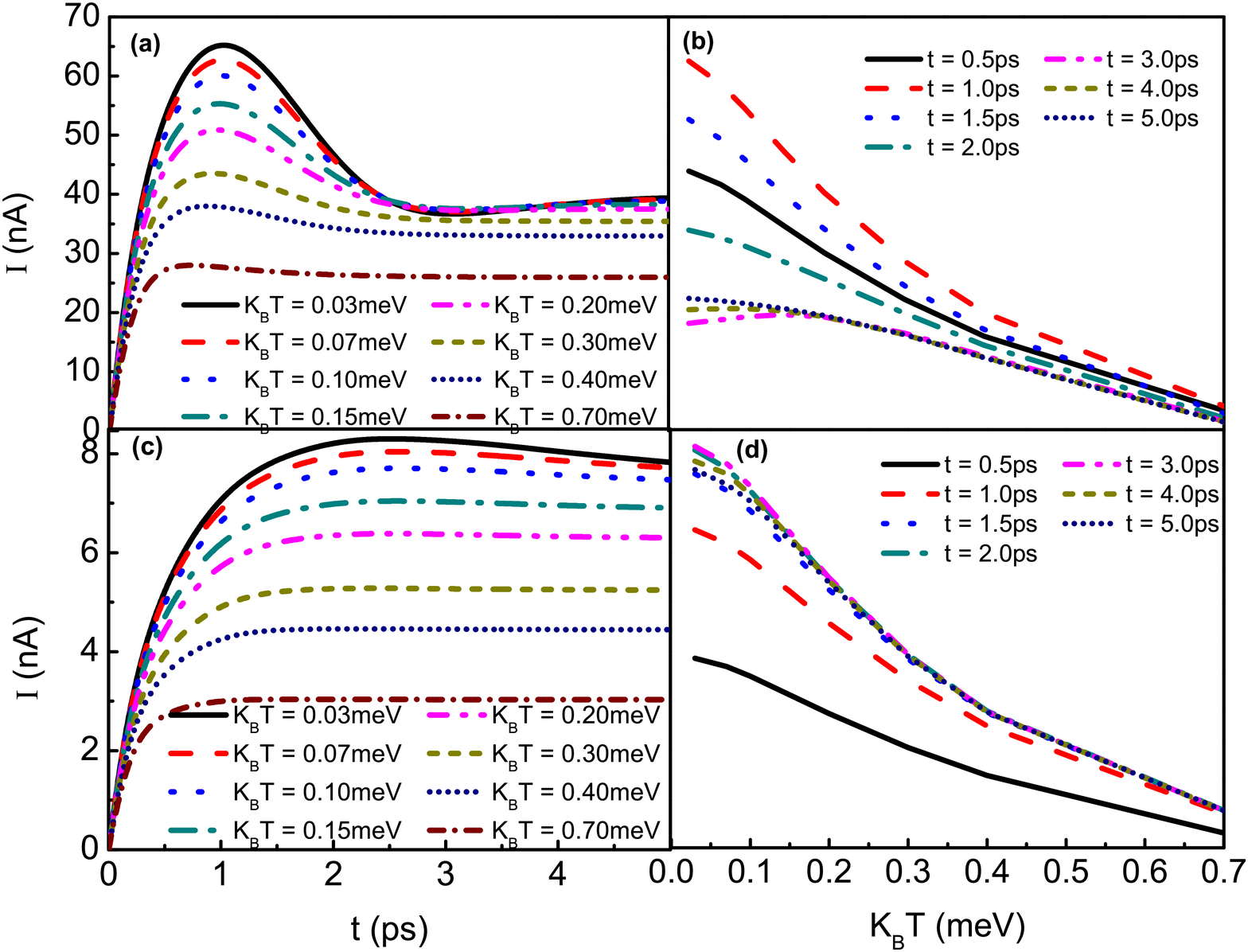}
\caption{(Color online). (a) Current $I(t)$ as a function of time $t$ (a) with different temperature $K_{B}T$ at a larger voltage $V_{SD} = 1.0\mathrm{meV}$. (b) The temperature $K_{B}T$ dependent of current $I(t)$ at times $t = 0.5 -5 \mathrm{ps}$ for a  larger voltage $V_{SD} = 1.0\mathrm{meV}$. (c) Current $I(t)$ as a function of time $t$ (a) with different temperature $K_{B}T$ at a small voltage $V_{SD} = 0.1\mathrm{meV}$. (d) The temperature $K_{B}T$ dependent of current $I(t)$ at times $t = 0.5 -5 \mathrm{ps}$ for a small voltage $V_{SD} = 0.1\mathrm{meV}$. The other parameters are $\varepsilon_{g} = 0\mathrm{meV}$, $U = 3\mathrm{meV}$, $W = 5\mathrm{meV}$, $\Delta = 0.3\mathrm{meV}$.}
\label{fig3}
\end{figure}

To check the details of the influence of temperature on the transient behavior of current $I(t)$, we alter the temperature in a widely ranging from $T \gg T_K$ to $T \ll T_K$, and present the results for the time dependent transport currents in figure 3. Figure 3 (a) shows the current $I(t)$ as a function of time $t$ with different temperatures $K_{B}T$ for a larger voltage $V_{SD} = 1.0\mathrm{meV}$. It is clearly visible that the nonlinear behavior is distinct with a large oscillating amplitude at low temperature (such as $K_{B}T = 0.03\mathrm{meV} $ curve). With the increase of temperature, the nonlinear response behavior will be suppressed gradually. Causing the amplitude of current to approach the steady state value. For example, the dynamic current changes to a linear response behavior at $K_{B}T = 0.70\mathrm{meV}$ and the amplitude of current only reaches the steady state value $I = 25\mathrm{nA}$. This indicates that the temperature plays an very important role on the nonlinear response behavior of current. At low temperature ($K_{B}T < 0.30\mathrm{meV}$), the time dependence of transport current is nonmonotonic. The current value anomaly first enhance and then decrease with time scale. At high temperature ($K_{B}T > 0.30\mathrm{meV}$), the QD system will fleetly reach the nonequilibrium steady state associated with a stable value of the current scaling with time (see figure \ref{fig3}(b)). This fast equilibration time results from the Kondo temperature of this mixed valence system \cite{2016prl036801}.

For comparison, we also sketches the temperature dependent current transition at a small bias voltage $V_{SD} = 0.1\mathrm{meV}$ in figure 3 (c) and (d). The nonlinear response behavior of current can not be realized for the widely temperature scale from $K_BT = 0.03\mathrm{meV}$ to $K_{B}T = 0.70\mathrm{meV}$. When the bias voltage pulse applied to the system, the current flowing through the device monotonically increase and reach a steady state value at the large times. The steady state value decreases with the high temperature. It can also be embodied in figure \ref{fig3}(d). For the small times ($t = 0.5\mathrm{ps}$ and $t = 1.0\mathrm{ps}$ curves), the current through the system sustains in linear increasing behavior which decreases with the high temperature. For the large times ($t > 1.5\mathrm{ps}$), the system will reach the nonequilibrium steady state with a constant value of current. Therefore, the temperature plays a restraining role on the dynamic transport current of QD system. Both the nonlinear behavior and steady state value of the current through the single QD system reveal continuously decreasing with the temperature increase.
\begin{figure}
\includegraphics[width=0.95\columnwidth]{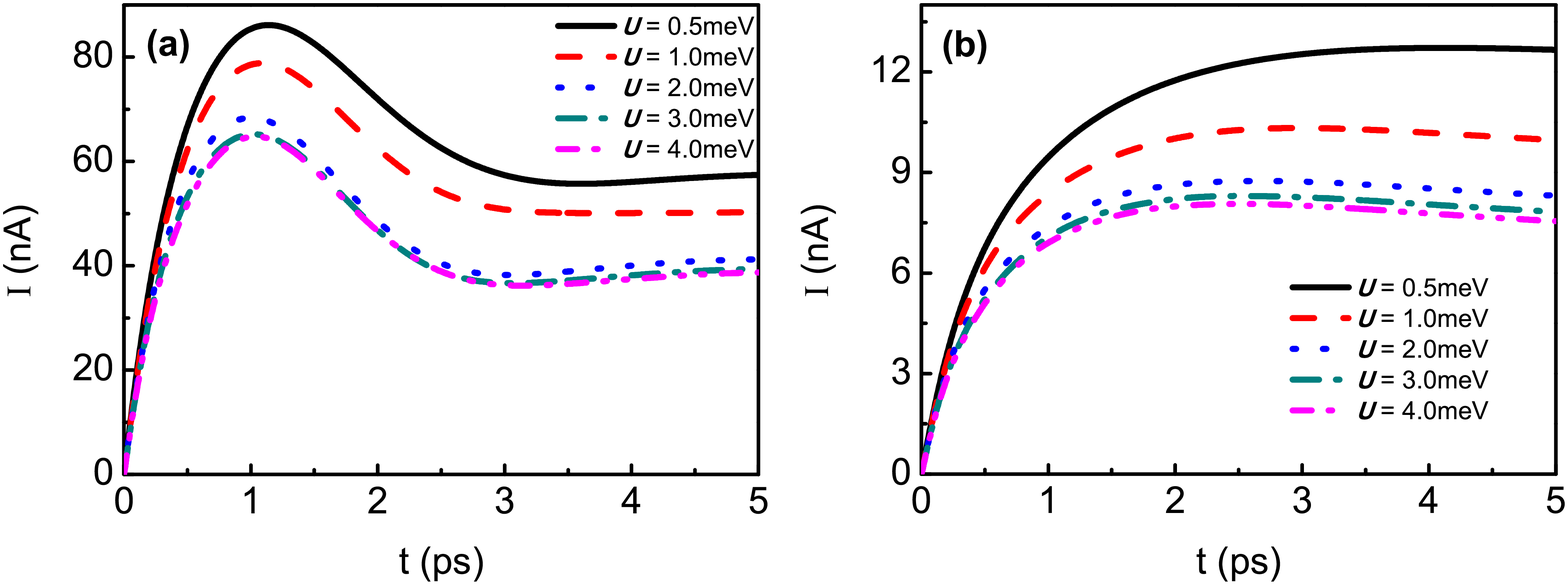}
\caption{(Color online). The $I(t)$-$t$ curves of the single quantum dot system with different \emph{e-e} interaction $U$ for a high voltage $V_{SD} = 1.0\mathrm{meV}$ (a) and for a low voltage $V_{SD} = 0.1\mathrm{meV}$ (b). The other parameters are $\varepsilon_{g} = 0\mathrm{meV}$, $W = 5\mathrm{meV}$, $\Delta = 0.3\mathrm{meV}$ and $K_{B}T = 0.03\mathrm{meV}$.}
\label{fig4}
\end{figure}

We then elucidate the influence of the finite strongly correlated electron-electron interaction \emph{U} on the transient dynamical behavior of the current, which is hard to treated by the other numerical approaches (such as TD-DMRG). The characteristics of transient dynamic $I(t)$-$t$ corresponding to different electron-electron interaction \emph{U} at larger voltage $V_{SD} = 1.0\mathrm{meV}$ and small voltage $V_{SD} = 0.1\mathrm{meV}$ are shown respectively in figure \ref{fig4} (a) and (b). It can be seen that the larger \emph{U} principally restrain the nonlinear behavior. With increasing the electron-electron interaction \emph{U}, the amplitude of the current oscillation decreases as well as a low value of steady state current. For example, the amplitude of current decreases from $85\mathrm{nA}$ at $U = 0.5\mathrm{meV}$ to $63\mathrm{nA}$ at $U = 2\mathrm{meV}$ as shown in figure \ref{fig4} (a). However, the engendering time of the nonlinear behavior is almost independent of the electron-electron interaction \emph{U}. As we known, the strongly correlated electron-electron interaction \emph{U} of the single QD corresponds to a transition of electrons from $n_d=1$ to $n_d=2$. A stronger \emph{U} will induce a larger distance between the two charge transfer peak associated with the spectrum of the single QD device which are located at $\varepsilon_{g}$ and $\varepsilon_{g}+U$ (see figure \ref{fig1} (b)). For the constant bias voltage, a stronger \emph{U} leads little density of states fall into the bias window. A low transient current flows through the single QD system associated with a small steady state value (see figure \ref{fig4} (b)).  As a consequence, the nonlinear behavior of the current will take more dramatically for small electron-electron interaction \emph{U} and the engendering time of the nonlinear behavior is independent of \emph{U}.
\begin{figure}
\includegraphics[width=0.95\columnwidth]{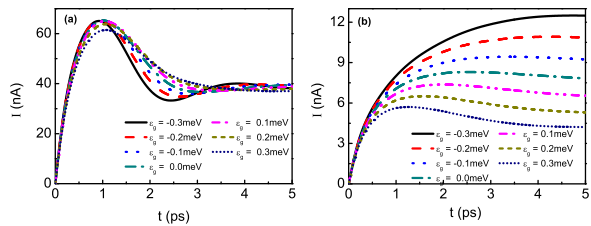}
\caption{(Color online). Current $I(t)$ as a function of time $t$ for different energy level of QD $\varepsilon_{g}$ for a low voltage $V_{SD} = 0.1\mathrm{meV}$ (a) and for a high voltage $V_{SD} = 1.0\mathrm{meV}$ (b). The other parameters are
$U = 3\mathrm{meV}$, $W = 5\mathrm{meV}$, $\Delta = 0.3\mathrm{meV}$ and $K_{B}T = 0.03\mathrm{meV}$.}
\label{fig5}
\end{figure}

Finally, to further study systematically the transient dynamical behavior of the current in the whole mixed valence regime, we alter the energy level of QD as $\mid \varepsilon_{g} \mid \leq \Delta$, which can be manipulated by the gate voltage $V_{g}$, and sketch the transition of the time dependent transport current through the single QD system in figure~\ref{fig5}. Figure~\ref{fig5} (a) and (b) depict the dynamical current $I(t)$ as a function of times $t$ after an high step voltage quench $V_{SD} = 1.0\mathrm{meV}$ and a low one $V_{SD} = 0.1\mathrm{meV}$, respectively. As shown in the figure, the high step voltage quench leads to extremely rich and various transport behaviors. Under the high step voltage quench condition ($V_{SD} = 1.0\mathrm{meV}$), the linear response behavior is embodied in the small times ($t < 1.0\mathrm{ps}$). After increasing to a maximal value, the transport current exhibits a nonlinear behavior for a larger time scale ($1.0\mathrm{ps} < t < 4.5\mathrm{ps}$). Finally, the current will get to a same steady state value for all energy levels of QD $\varepsilon_{g}$. With the energy level of QD elevating, the nonlinear behaviors of the current are strangely subdued accompanied by a low oscillating amplitude both for occupied state ($\varepsilon_{g} < 0$) and  unoccupied state ($\varepsilon_{g} > 0$) (see figure ~\ref{fig5} (a)). For the low step voltage quench case ($V_{SD} = 0.1\mathrm{meV}$), the transform of the current is ordinary. When the energy of the localized state is below the Fermi level as the occupied state ($\varepsilon_{g} < 0$), the current decrease with the higher energy level of QD. And for the unoccupied state ($\varepsilon_{g} > 0$), the non-linear behavior emerges at $\varepsilon_{g} > 0.2\mathrm{meV}$ as shown in figure \ref{fig5} (b). Here, the empty orbit of single QD device leads the temporal coherence of electrons tunneling. More over, the steady state values of the current decrease with elevating the energy level of QD $\varepsilon_{g}$  for the low step voltages.

\section{CONCLUSIONS}

In summary, we have investigated the transient dynamics properties response of strongly correlated QD system in the mixed valence regime based on the  hierarchical equations of motion. The time-dependent transport current and occupations in a range of temperature below and above the Kondo temperature are explored. The current shows a monotonic rising behavior for the low bias voltages and an nonlinear behavior for the strong bias voltages which can be demonstrated by the transition of the voltage dependent QD occupancies. The temperature plays an very important role on the nonlinear response behavior of current. At low temperature case, the time dependence of transport current is nonmonotonic with a distinct nonlinear behavior. The current value anomaly first enhance and then decrease with time scale. At high temperature case, the nonlinear response behavior will be suppressed. The QD system will fleetly reach the nonequilibrium steady state with a stable value of the current scaling with times. We also find that the nonlinear behavior of the current will take more dramatically for small electron-electron interaction and the engendering time of the nonlinear behavior is independent of $U$. The energy level of QD in the mixed valence regime can also lead the coexistence nonlinear behaviors of the transient current. The nonlinear behaviors of the current in the occupied state are more extreme than unoccupied state. With the energy level of QD elevating, the nonlinear behaviors of the current are strangely subdued accompanied by a low oscillating amplitude both for occupied state and unoccupied state. Those characteristics may be observed in experiments.

\acknowledgments
This work was supported by the NSF of China (No.\,11374363) and the Research Funds of Renmin University of China (Grant No. 11XNJ026). Computational resources have been provided by the
Physical Laboratory of High Performance Computing at Renmin University of China.

\end{document}